\documentclass[usenatbib,iop,numberedappendix]{aeb_emulateapj_2010}
\usepackage{amsmath}
\usepackage{amssymb}

\usepackage{natbib}

\usepackage[usenames,dvips]{color}

\def\s{{\rm s}} 
\def\Ms{{\rm M}\s} 

\def\GHz{{\rm GHz}} 

\def\m{{\rm m}} 
\def\mm{{\rm m}\m} 
\def\km{{\rm k}\m} 

\def\Ms{M_\odot} 

\def\Jy{{\rm Jy}} 

\def\muas{\mu{\rm as}} 

\renewcommand{\d}{{d}}
\newcommand{\e}{{e}}
\def\Ms{{M_\odot}}
\def\Rs{{r_{\rm S}}}

\newcommand\bmath[1] {\mbox{\boldmath$\rm #1$}}

\begin{document}

\title{
Constraining the Structure of Sagittarius A*'s Accretion Flow\\
with Millimeter-VLBI Closure Phases
}

\author{
Avery E.~Broderick\altaffilmark{1},~
Vincent L.~Fish\altaffilmark{2},~
Sheperd S.~Doeleman\altaffilmark{2} and
Abraham Loeb\altaffilmark{3}
}
\altaffiltext{1}{Canadian Institute for Theoretical Astrophysics, 60 St.~George St., Toronto, ON M5S 3H8, Canada; aeb@cita.utoronto.ca}
\altaffiltext{2}{Massachusetts Institute of Technology, Haystack Observatory, Route 40, Westford, MA 01886.}
\altaffiltext{3}{Institute for Theory and Computation, Harvard University, Center for Astrophysics, 60 Garden St., Cambridge, MA 02138.}

\shorttitle{Sgr A*'s Closure Phases}
\shortauthors{Broderick et al.}

\begin{abstract}
Millimeter wave Very Long Baseline Interferometry (mm-VLBI) provides access
to the emission region surrounding Sagittarius A*, the supermassive
black hole at the center of the Milky Way, on sub-horizon scales.
Recently, a closure phase of $0^\circ\pm40^\circ$ was reported on a
triangle of Earth-sized baselines (SMT-CARMA-JCMT) representing a new
constraint upon the structure and orientation of the emission region,
independent from those provided by the previously measured
$1.3\mm$-VLBI visibility amplitudes alone. Here, we compare this to
the closure phases associated with a class of physically motivated,
radiatively inefficient accretion flow models, and present predictions
for future mm-VLBI experiments with the developing {\em Event Horizon
  Telescope} (EHT).  We find that the accretion flow models are
capable of producing a wide variety of closure phases on the
SMT-CARMA-JCMT triangle, and thus
not all models are
consistent with the recent
observations.  However, those models that reproduce the
$1.3\mm$-VLBI visibility amplitudes overwhelmingly have SMT-CARMA-JCMT
closure phases between $\pm30^\circ$, and are therefore broadly
consistent with all current mm-VLBI observations.  Improving station
sensitivity by factors of a few, achievable by increases in bandwidth
and phasing together multiple antennas at individual sites, should
result in physically relevant additional constraints upon the model
parameters and eliminate the current $180^\circ$ ambiguity on the
source orientation.  When additional stations are included, closure
phases of order $45^\circ$--$90^\circ$ are typical.
In all cases the EHT will be able to measure these with sufficient
precision to produce dramatic improvements in the constraints upon the
spin of Sgr A*.
\end{abstract}


\maketitle

\section{Introduction} \label{I}
It has recently become possible to study the emission regions of a
handful of black holes on sub-horizon scales with millimeter wave Very Long
Baseline Interferometry (mm-VLBI). 
Already, this
technique has produced horizon-scale information on Sagittarius A* (Sgr A*), the
$4.3\times10^6\,\Ms$ black hole located at the center of the Milky
Way \citep{Ghez_etal:08,Gill_etal:09a,Gill_etal:09b}, using the
{\em Arizona Radio Observatory Sub-Millimeter Telescope} (SMT) on Mount Graham, Arizona, 
{\em James Clerk Maxwell Telescope} (JCMT) and
{\em Sub-Millimeter Array} (SMA) atop Mauna Kea in Hawaii, and the
{\em Combined Array for Research in Millimeter-wave Astronomy} (CARMA) in Cedar Flat,
California \citep{Doel_etal:08,Fish_etal:10}.
Due to the limited signal-to-noise of these early experiments, they have
produced primarily visibility amplitudes, related to the modulus of
the Fourier transform of the intensity distribution of the source.
Nevertheless, when analyzed in the
context of physically motivated accretion flow models, they have
resulted in dramatic constraints upon the spin orientation and
magnitude \citep{Brod_etal:09,Huan-Taka-Shen:09,Dext-Agol-Frag-McKi:10,Brod_etal:10}.
However, the absence of phase information introduces fundamental
degeneracies in the orientation of the modeled image, and systematic
uncertainties in the image structure generally.

The importance of visibility phase information has been appreciated
since the beginning of radio interferometry.  In the context of VLBI,
it has only been widely possible since the introduction of ``closure
phases'', and the associated development of self-calibration
techniques
\citep[e.g.,][]{Jenn:58,Jenn-Lath:59,Roge_etal:74,Pear-Read:84,Thom-Mora-Swen:01}.
The closure phases, which are equivalent to the argument of the bispectrum, are
combinations of the visibility phases on  baseline triangles (and thus
VLBI station triples) that are insensitive to individual station-based phase
errors, which otherwise typically dominate the phase
uncertainties, e.g., due to pathlength variations from atmospheric turbulence.\footnote{Due to the degeneracy amongst possible
  triangles, given $N$ antennas there are only $(N-1)(N-2)/2$
  independent closure phases, and thus $N-1$ additional phases must be
  supplied to produce the full compliment of phase information.
  Frequently, these are obtained via self-calibration techniques, a
  combination of non-linear algorithms in which the unknown phases are
  chosen such that the resulting image satisfies various physical
  constraints \citep[see, e.g.,][for a detailed summary of
  self-calibration techniques]{Pear-Read:84}.}
Specifically, if $\phi_{ij}$ is the visibility phase on the baseline
between stations $i$ and $j$, the closure phase associated with three
stations is
\begin{equation}
\Phi_{ijk}\equiv\phi_{ij}+\phi_{jk}+\phi_{ki}\,.
\end{equation}
Even a handful of closure phases are diagnostic of the underlying
image structure, e.g., the closure phases of a point source or
Gaussian flux distribution (including asymmetric cases) are
identically $0^\circ$, while that of an annulus may be either $0^\circ$ or
$180^\circ$, depending upon the particular baselines considered.  More
general flux distributions produce non-vanishing closure phases,
indicative of the symmetry of the image.
Recently, \citet{Fish_etal:10} reported the measurement of a closure
phase at $1.3\,\mm$ for Sgr A* of $0^\circ\pm40^\circ$ on the
SMT-JCMT-CARMA triangle.  Note that this represents a new constraint
upon models for the structure of the emitting region surrounding Sgr
A*, independent of those associated with the visibility amplitudes
alone.

The {\em Event Horizon Telescope} (EHT) is a project underway that
will extend current mm-VLBI arrays to shorter wavelengths
($0.8\,\mm$), increased sensitivity, and greater baseline coverage,
substantially improving the ability of mm-VLBI to study black holes
on Schwarzschild radius scales.
\citep{Doel_etal:2010}.  Anticipated and
potential future stations sites include Chile ({\em Atacama Pathfinder EXperiment},
{\em Atacama Submillimeter Telescope} and {\em Atacama Large Millimeter Array};
APEX, ASTE, and ALMA, respectively), Mexico ({\em Large Millimeter
Telescope}; LMT), the South Pole ({\em South Pole Telescope}; SPT),
and the IRAM telescopes in Spain ({\em Pico Veleta}; PV) and France
({\em Plateau de Bure}; PdB).  Among these the longest baselines are
$u\simeq1.2\times10^4\,\km$, corresponding to a maximum angular
resolution of $\lambda/2u\simeq11\,\muas$ at $230\,\GHz$ ($1.3\,\mm$)
and $7.5\,\muas$ at $345\,\GHz$ ($0.87\,\mm$).  Expected increases in
bandwidth and the phasing together of elements within Hawaii, Chile
and CARMA \citep[e.g.,][]{Wein:08}, will lead to substantial improvements in
sensitivity.  As a consequence, it will become possible in the near
future to measure mm-VLBI closure phases on a variety of additional
triangles with uncertainties considerably smaller than that of the
\citet{Fish_etal:10} result.

Here, motivated by previous efforts to model the $1.3\,\mm$-VLBI
visibilities using physically motivated accretion models for Sgr A*,
we compute the closure phases implied by the radiatively inefficient
accretion flow models and $1.3\,\mm$-VLBI visibility amplitude fits
presented in \citet{Brod_etal:10}.  By doing so we address three
immediate questions:
\begin{enumerate}
\item Is the new closure phase estimate consistent with the accretion
  flow models we have considered in particular and radiatively
  inefficient accretion flow models generally?
\item What is the strength of the constraint placed upon physically
  motivated accretion flow models and the estimates of black hole spin
  by the measured closure phase?
\item What are the strength of the constraints that will possible in
  the near future as the EHT develops?
\end{enumerate}

In Section \ref{sec:CAFCP} we briefly describe the accretion models
and how we compute the closure phases.  In Section \ref{sec:CwMV} we
compare the predicted closure phases with the measured values.  In
Section \ref{sec:CPFT} we predict the closure phases for the EHT and
compare these with the estimated uncertainties of the EHT.  Finally,
we summarize our conclusions in Section \ref{sec:C}.

\section{Computing Accretion Flow Closure Phases} \label{sec:CAFCP}

\subsection{Accretion Modeling}
We model Sgr A*'s accretion flow as a radiatively inefficient
accretion flow, the details of which may be found in
\citet{Brod_etal:10}, and references therein, and thus we only
summarize the model here.

Sgr A* transitions from an inverted, presumably optically thick
spectrum to an optically thin spectrum near millimeter wavelengths.
This implies that near 1.3mm Sgr A* is only becoming optically thin,
and thus absorption in the surrounding medium is likely to be
important.  This transition does not occur isotropically, happening at
longer wavelengths for gas that is receding and at shorter wavelengths
for gas that is approaching.  Therefore, properly modeling the
structure and relativistic radiative transfer is crucial to producing
high fidelity images.

For concreteness, as in \citet{Brod_etal:10}, we follow
\citet{Yuan-Quat-Nara:03} and employ a model in which the accretion
flow has a Keplerian velocity distribution, a population of thermal
electrons with density and temperature
\begin{equation}
n_{e,{\rm th}}=n_{e,{\rm th}}^0 \left(\frac{r}{\Rs}\right)^{-1.1} e^{-z^2/2\rho^2}
\end{equation}
and
\begin{equation}
T_{e}=n_{e}^0 \left(\frac{r}{\Rs}\right)^{-0.84}\,,
\end{equation}
respectively,
and a toroidal magnetic field in approximate ($\beta=10$)
equipartition with the ions (which are responsible for the majority of
the pressure), i.e.,
\begin{equation}
\frac{B^2}{8\pi} = \beta^{-1} n_{e,{\rm th}} \frac{m_p c^2 \Rs}{12 r}\,.
\end{equation}
In all of these, $\Rs=2GM/c^2$ is the Schwarzschild radius, $\rho$ is
the cylindrical radius and $z$ is the vertical coordinate.  Inside of
the innermost-stable circular orbit (ISCO) we assume the gas is
plunging upon ballistic trajectories.  In principle the plunging gas
can still radiate, though in practice it contributes little to the
overall emission due to the large radial velocities it develops.  In
the case of the thermal quantities the radial structure was taken from
\citet{Yuan-Quat-Nara:03}, and the vertical structure was determined
by assuming that the disk height is comparable to $\rho$.  Note that
all of the models we employ necessarily have the spin aligned with the
orbital angular momentum of the accretion flow.  For the regions that
dominate the $\mm$ emission, this assumption is well justified due to
disk precession and viscous torques, though it may be violated at
large distances.

Thermal electrons alone are incapable of reproducing the nearly-flat
spectrum of Sgr A* below $43\,\GHz$.  Thus it is necessary to also
include a nonthermal component.  As with the thermal components, we
adopt a self-similar model for a population of nonthermal electrons,
\begin{equation}
n_{e,{\rm nth}}=n_{e,{\rm nth}}^0 \left(\frac{r}{\Rs}\right)^{-2.02} e^{-z^2/2\rho^2}\,,
\end{equation}
with a power-law distribution corresponding to a spectral index of
$1.25$ and cut off below Lorentz factors of $10^2$ \citep[consistent with
][]{Yuan-Quat-Nara:03}.  The radial power-law index was chosen to
reproduce the low frequency spectrum of Sgr A*, and is insensitive to
the black hole properties due to the distant location of the
long-wavelength emission.

The primary emission mechanism at the wavelengths of interest is
synchrotron radiation, arising from both the thermal and nonthermal electrons.
We model the emission from the thermal electrons using the emissivity
described in \citet{Yuan-Quat-Nara:03}, appropriately altered to
account for relativistic effects \citet[see, e.g., ][]{Brod-Blan:04}.
Since we perform polarized radiative transfer via the entire
complement of Stokes parameters, we employ the polarization fraction
for thermal synchrotron as derived in \citet{Petr-McTi:83}.  In doing
so, we have implicitly assumed that the emission due to thermal
electrons is isotropic, which while generally not the case is unlikely
to change our results significantly.  For the nonthermal electrons, we
follow \citet{Jone-ODel:77} for a power-law electron distribution,
with an additional spectral break associated with the minimum electron
Lorentz factor.  For both emission components the absorption
coefficients are determined directly via Kirchhoff's law.  Images are
then produced using the fully relativistic ray-tracing and radiative
transfer schemes described in \citet{Brod-Loeb:06a,Brod-Loeb:06b} and
\citet{Brod:06}.

Because \citet{Yuan-Quat-Nara:03} neglected relativistic effects and
assumed spherical symmetry, it is not directly applicable here.  For
these reasons, as in \citet{Brod_etal:10}, the coefficients
$(n_{e,{\rm th}}^0,T_e^0,n_{e,{\rm nth}}^0)$ were adjusted to fit the
radio spectral energy distribution (SED) of Sgr A* \citep[see ][ for
  details on the fitting procedure]{Brod_etal:10}.  We repeated this
procedure for a large number of positions in the
dimensionless spin-inclination ($a$-$\theta$) parameter
space\footnote{The black hole angular momentum is given by
  $a GM^2/c$, with $a=1$ corresponding to a maximally rotating Kerr
  spacetime.  We define $\theta$ such that models with
  $\theta=0^\circ$ are viewed along the spin axis, and thus the thick
  accretion disk viewed face-on.  Conversely, models with
  $\theta=90^\circ$ are viewed perpendicular to the spin axis, and
  thus the thick accretion disk is viewed edge-on.},
producing a tabulated 
set of the coefficients $(n_{e,{\rm th}}^0,T_e^0,n_{e,{\rm nth}}^0)$ at a
large number of points
throughout the $a$-$\theta$ parameter space.  In all cases it was
possible to fit the SED with extraordinary precision.
From the
tabulated values, the coefficients are then obtained at arbitrary $a$
and $\theta$ using high-order polynomial interpolation.

During the $1.3\mm$-VLBI observations Sgr A*'s flux varied by roughly
$30\%$.  We model this as a variable accretion rate, moving the
electron density normalization up and down.  In practice, we reduced
the electron density normalization by an amount sufficient to produce
a total flux of $2.5\,\Jy$, and then multiplied the resulting images
by a correction faction during the mm-VLBI data analysis.  Because the
source is not uniformly optically thin, this is not strictly correct,
though this makes a small change to the images themselves.  We
produced 9090 images, with flux normalized as described above, at
$a\in\{0,0.01,0.02,...,0.98,0.99,0.998\}$ for each
$\theta\in\{1^\circ,2^\circ,...,89^\circ,90^\circ\}$.  We then produce
models with arbitrary position angles, $\xi$, by rotating the images on
the sky.  For this purpose we define $\xi$ as the position angle
(east of north) of the projected spin vector.

\subsection{Computing the Closure Phases}
The intrinsic complex visibilities are obtained in the standard fashion:
\begin{equation}
V(u,v) = \iint \d\alpha\,\d\beta \e^{-2\pi i(\alpha u +\beta v)/\lambda} I(\alpha,\beta)\,,
\end{equation}
where $I(\alpha,\beta)$ is the intensity at a given set of angular
coordinates.  These are subsequently modified to account for the
observed interstellar electron scattering
\citep[see, e.g.,][]{Bowe_etal:06,Brod_etal:10}\footnote{Note that
  the empirically measured Gaussian scattering kernel does not produce
phase shifts in the visibilities, and thus we may neglect the
scattering in computing the closure phases.  Nevertheless, we must keep
this effect for the computation of the expected closure phase
uncertainties.}.  The closure phase for a given triplet of
observatories is then given by
\begin{equation}
\Phi_{ijk}
=
\arg\left[V(u_{ij}, v_{ij})\right]
+\arg\left[V(u_{jk}, v_{jk})\right]
+\arg\left[V(u_{ki}, v_{ki})\right]\,.
\end{equation}

We compute the $\Phi_{ijk}$ for all spin magnitude, inclination and
position angles considered in \citet{Brod_etal:10}.  From these we
determine the probability density of a given $\Phi_{ijk}$:
\begin{equation}
p(\Phi) = \int \d^3\!a \,p(\bmath{a}) \,\delta\left[ \Phi_{ijk}(\bmath{a})-\Phi \right]\,,
\end{equation}
in which $p(\bmath{a})$ is the probability of a given vector spin.
For our purposes here we consider two forms of $p(\bmath{a})$: an
isotropically distributed spin direction with a flat prior upon the
spin magnitude, the same prior as that adopted in
\citet{Brod_etal:10}, and the posterior probability of a given spin,
defined in \citet{Brod_etal:10}, after fitting the $1.3\mm$-VLBI
amplitudes.  We will refer to these as the ``isotropic'' and
``amplitude-fitted'' priors, respectively. 
Interpolation errors associated with the finite density of points in
the spin parameter space and the finite size of the imaged region,
resulted in some small-amplitude  saw-tooth features in the $p(\Phi)$.  As a
consequence, we smoothed the $p(\Phi)$ shown on scales of
$1^\circ$--$9^\circ$, depending upon the particular case, with
$5^\circ$ being typical.  Where large spike features appeared we did
no smoothing, and in all cases the wings of the probability
distribution remained unchanged.

Note that fitting the visibility amplitudes necessarily imparts a
$180^\circ$ degeneracy in the position angle, corresponding to a sign
change in the resulting closure phase.  As a result, in principle, all
closure phase estimates we present are unique only up to a sign.
Nevertheless, we show the closure phases associated with one
particular choice for the position angle, specifically that for which
the spin of the most likely configuration is oriented $-52^\circ$ east
of north.

\begin{figure}
\begin{center}
\includegraphics[width=0.5\textwidth]{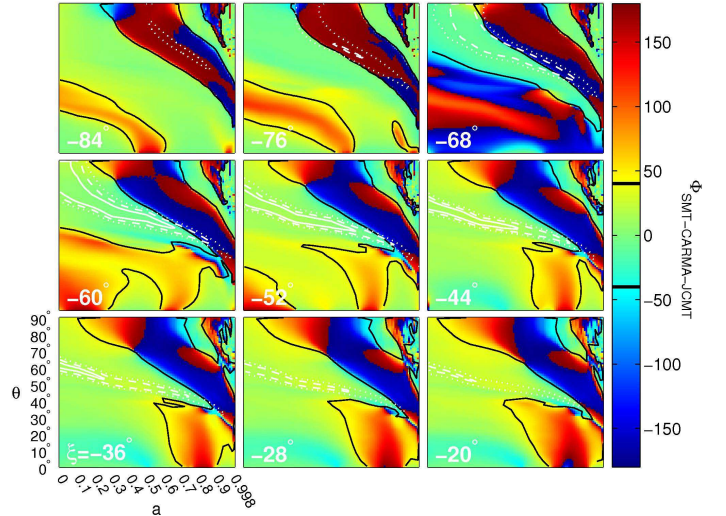}
\end{center}
\caption{SMT-CARMA-JCMT closure phases, determined at 12:20 UTC on Day
  96 of 2009, for the accretion models considered by
  \citet{Brod_etal:10} as a function of dimensionless spin magnitude
  ($a$) and inclination ($\theta$) for a number of position angles ($\xi$). 
  Black contour lines show the closure phase limit of $0^\circ\pm40^\circ$
  reported by \citet{Fish_etal:10}.  For
  reference, the posterior probability contours (white) associated with the
  $1$--$\sigma$ (solid), $2$--$\sigma$ (dashed), and $3$--$\sigma$
  (dotted) cumulative probabilities as determined in
  \citet{Brod_etal:10} are overlaid. 
  Note that these are degenerate to
  $\Phi\rightarrow-\Phi$.}\label{fig:1220CPs}
\end{figure}
\begin{figure}
\begin{center}
\includegraphics[width=0.5\textwidth]{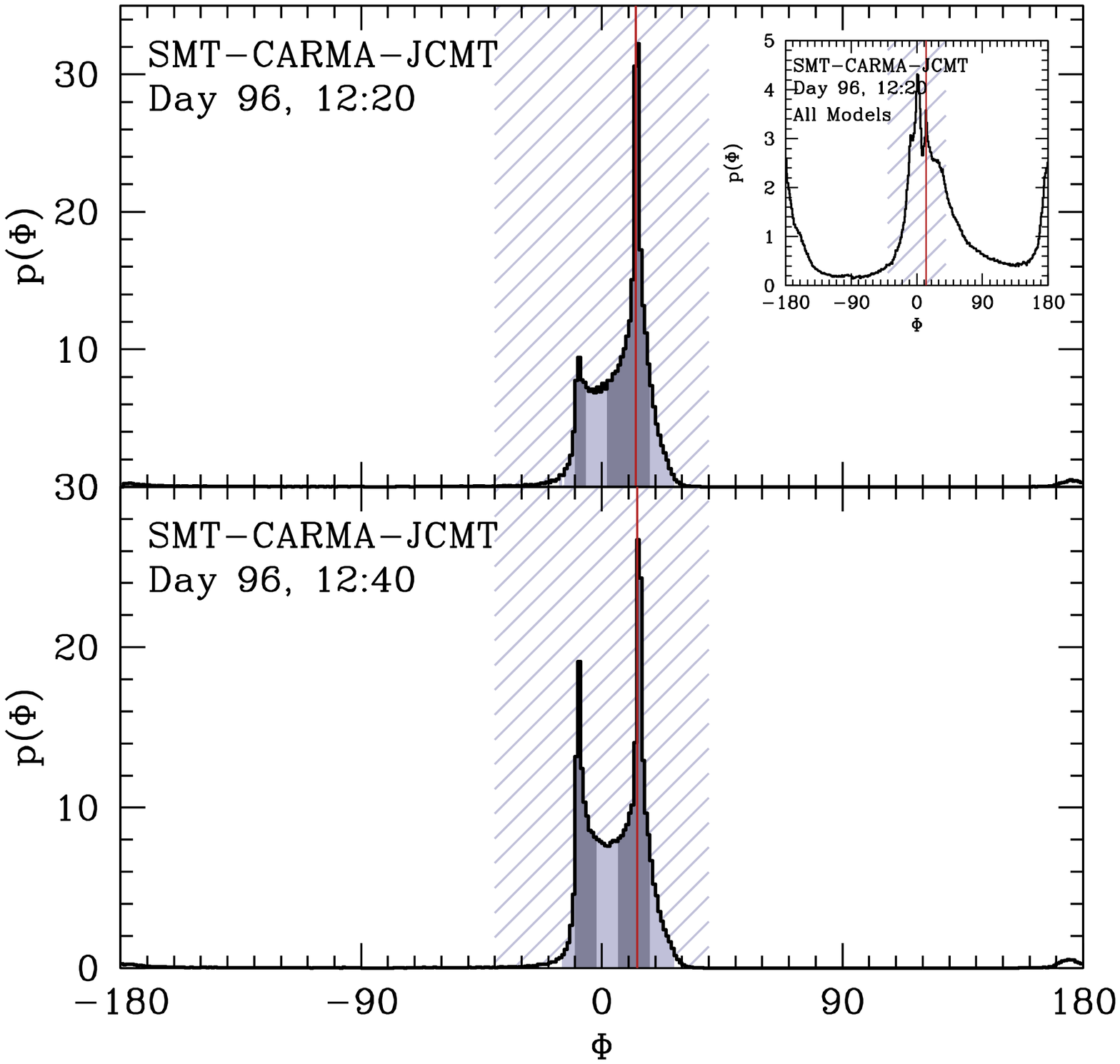}
\end{center}
\caption{Probability distribution of the closure phases associated
  with the accretion flow models considered by \citet{Brod_etal:10}
  for the SMT-JCMT-CARMA triangle during the times at which
  \citet{Fish_etal:10} report measurements of this quantity.  The dark
  grey and light grey regions are the $1$--$\sigma$ and $2$--$\sigma$
  regions, defined by cumulative probability, and the light grey
  hatched region shows the measured value ($0^\circ\pm40^\circ$).  The
  vertical red line indicates the closure phase associated with the
  most likely accretion flow model.  For reference, the distribution
  of closure phases when the $1.3\mm$-VLBI prior is not applied is
  shown in the upper inset.  Note that these are degenerate to
  $\Phi\rightarrow-\Phi$.}\label{fig:Day96CPs}
\end{figure}

\subsection{Closure Phase Uncertainties}
\begin{deluxetable}{lrcc}
\tablecaption{\mbox{Estimated System Equivalent Flux Densities at $230\,\GHz$\label{tab:SEFDs}}}
\tablehead{
\colhead{Facility~~~~~~~~~~~~} &
\colhead{~~~~~~~$N$\tablenotemark{a}} &
\colhead{~~~~~~~Diameter\tablenotemark{a} (m)} &
\colhead{~~~~~~~SEFD \tablenotemark{b} (Jy)}\\
}
\startdata
Hawaii & 8 & 23 & 4900\\
CARMA  & 8 & 27 & 6500\tablenotemark{c}\\
SMT    & 1 & 10 & 11,900\\
LMT    & 1 & 32 & 10,000\tablenotemark{d}\\
APEX   & 1 & 12 & 6500\\
ALMA   & 10 & 38 & 500\\
\enddata
\tablenotetext{a}{Effective aperture when number of antennas ($N$) are
  phased together.}
\tablenotetext{b}{Expected system equivalent flux density values
  toward Sgr A* include typical weather conditions and opacities.}
\tablenotetext{c}{Based upon recent observations.}
\tablenotetext{d}{Completion of the dish and upgrades to the surface
  accuracy and receiver will eventually lower the SEFD by more than a
  factor of 10.}
\end{deluxetable}
The precision with which a given closure phase may be measured depends
upon the signal-to-noise ratio (SNR) of the
visibilities on the individual baselines,
$s_{ij}\equiv V_{ij}\sqrt{2B\tau/{\rm SEFD}_i{\rm SEFD}_j}$, where
$V_{ij}$ is the visibility amplitude, $B$ is the bandwidth, $\tau$ is
the atmospheric coherence time, and ${\rm SEFD}_i$ is the system
equivalent flux density for a given station.
Here we take $\tau=10\,\s$, typical of the atmospheric coherence times at
the wavelengths and sites of interest.  However, because closure phases are
insensitive to atmospheric phase fluctuations, they may be
coherently averaged over time scales much larger than $\tau$, with the
measurement uncertainties decreasing as $\sqrt{\tau/T}$.  This is
limited by the time scales on which the baseline orientations and the
intrinsic structure of Sgr A* change significantly.  In quiescence,
the former is the limiting factor, and thus we choose $T=10\,\min$.
However, during periods of flaring activity, or if Sgr A* exhibits
some dynamical
structure, this could be considerably shorter.  Finally, unless
otherwise stated, we assume $B=4\,\GHz$, the anticipated near-term
bandwidth target of the EHT, and adopt the ${\rm SEFD}_i$ reported in
Table 1 of \citet{Doel_etal:09}, partially reproduced in Table
\ref{tab:SEFDs} here.\footnote{This choice of bandwidth and set of
  ${\rm SEFD}_i$ results in baseline SNRs approximately an order of
  magnitude larger than those obtained during the $1.3\mm$-VLBI
  observations reported in \citet{Fish_etal:10}.}
Specifically we assume that the SMA, CSO and JCMT are phased together
in Hawaii, six of the $10.4\,\m$
and two of the $6\,\m$ CARMA antennas are phased together, and
ten of the $12\,\m$ ALMA antennas are phased together in Chile.
In terms of the $s_{ij}$, the closure phase uncertainty is given by,
\begin{equation}
\sigma_{\Phi_{ijk}}
=
\sqrt{\frac{\tau}{T}} L^{-1}\left[ L(s_{ij})L(s_{jk})L(s_{ki}) \right]\,,
\end{equation}
where
\begin{equation}
L(s) \equiv \sqrt{\frac{\pi}{8}} s e^{-s^2/4} \left[
  I_0\left(\frac{s^2}{4}\right)+I_1\left(\frac{s^2}{4}\right)
\right]
\end{equation}
in which $I_0$ and $I_1$ are the hyperbolic Bessel functions of the
first kind \citep{Roge_etal:84,Roge-Doel-Mora:95}.  In the high SNR limit
($s_{ij},\,s_{jk},\,s_{ki}\gg1$) this reduces to
\begin{equation}
\sigma_{\Phi_{ijk}}
\simeq
\sqrt{\frac{\tau}{T}} \left[ s_{ij}^{-2} + s_{jk}^{-2} + s_{ki}^{-2} \right]^{-1/2}\,,
\end{equation}
though we will make use of the exact expression.  Note that since
$\sigma_\Phi$ depends upon the individual SNR values, it is model
dependent; the values we quote are associated with the most likely
accretion model, as determined via the $1.3\,\mm$-VLBI visibility
amplitudes alone.

\section{Closure Phase on the SMT-CARMA-JCMT Triangle: Comparison with Measured values} \label{sec:CwMV}
\begin{figure*}
\begin{center}
\includegraphics[width=\textwidth]{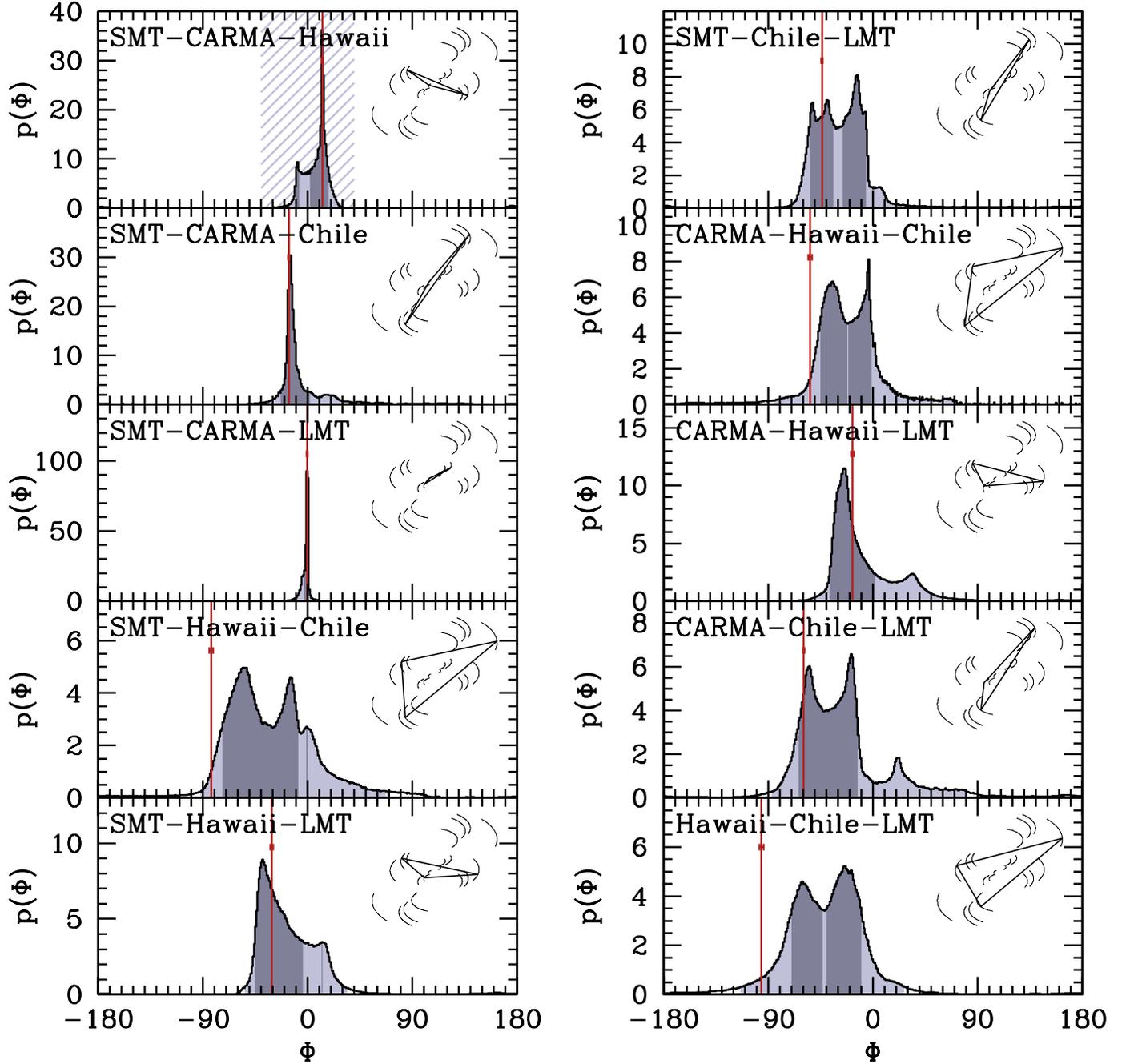}
\end{center}
\caption{Probability distributions for the closure phases at 12:20 UTC
  on day 96 of 2009 associated
  with the accretion flow models considered by \citet{Brod_etal:10}
  for all triangles constructed from the SMT, Hawaii, CARMA, LMT
  and Chile stations.  The hatched region in the upper left indicates the
  closure phase measurement reported in \citet{Fish_etal:10}.  In all
  plots the dark and light grey regions are the $1$--$\sigma$ and
  $2$--$\sigma$ regions, defined by cumulative probability, the
  vertical red lines show the closure phase of the most likely
  accretion flow model, and the red horizontal error bar indicates the
  accuracy with which the EHT should be able to measure the closure
  phases in the future on $10\,\min$ time scales (using 10 ALMA antennas
  in Chile, employing the recently phased sites in Hawaii, and 8 CARMA
  antennas).  For reference the
  $u$--$v$ tracks are shown in the upper right of each plot, with the
  triangle for the closure phase plotted indicated explicitly.  Note
  that these are degenerate to $\Phi\rightarrow-\Phi$.}\label{fig:AllCPs}
\end{figure*}
There are two $15\,\min$ periods over which Sgr A*'s closure phase was
measured on the SMT-JCMT-CARMA triangle on the night of April 6, 2009
(corresponding to day 96), beginning on 12:20 UTC and 12:40 UTC
\citep{Fish_etal:10}.  The distribution of the closure phases at 12:20
UTC in spin magnitude ($a$), spin inclination ($\theta$), and spin
position angle ($\xi$) are shown in Figure \ref{fig:1220CPs}, with the
closure phases at 12:40 UTC differing by only a small amount.  For
reference the $1$--$\sigma$, $2$--$\sigma$, and $3$--$\sigma$
posterior probability contours
from the $1.3\,\mm$-VLBI visibility fits obtained in \citet{Brod_etal:10} are
overlaid, shown by the solid, dashed, and dotted white contours,
respectively.

It is immediately apparent that small closure phases are not generic.
Assuming the isotropic prior, fewer than 49\%
of the accretion flow models are consistent with the closure phase
measurement.  The inset in Figure \ref{fig:Day96CPs} bears this out;
while small closure phases are marginally preferred, a large closure
phase island exists with a substantial likelihood of any phase being
measured.  The excluded models are not confined to a particular region
in the spin parameter space, extending over a wide range of spin
magnitudes, inclinations and position angles.  Thus, despite the large
uncertainties, the measured closure phase has considerable diagnostic
ability.

However, this changes dramatically when we restrict our attention to
those models which are consistent with the $1.3\mm$-VLBI amplitudes.
The white contours in Figure \ref{fig:1220CPs} show the posterior
probability of a given spin vector based upon the $1.3\mm$-VLBI visibility
amplitudes alone, i.e., the amplitude-fitted prior.  The regions of
high probability are confined to a narrow valley of low closure
phases.  Hence, while accretion models generally are not necessarily
consistent with the observed closure phases, the vast majority of
those that were found to reproduce the visibility amplitudes {\em are}
broadly consistent with the recent measurement.  Once the
amplitude-fitted prior is adopted, the probability of finding a
closure phase that is larger than $40^\circ$ falls to less than 3\%.

The amplitude-fitted models that violate the recent closure phase
measurement are confined to the $2$--$\sigma$ significance regions,
primarily at large spin magnitude and large negative position angles,
where the closure phase is $\sim180^\circ$.  This behavior is similar
to that exhibited by annular models, in which the closure phase varies
between $0^\circ$ and $180^\circ$, depending upon the angular size of
the annulus.  As a result, improvements of factors of a few in
sensitivity will allow closure phase measurements that can compete
with the visibility amplitudes in their power to constrain the 
structure of Sgr A*.

These conclusions are borne out by the closure phase probability
distributions shown in Figure \ref{fig:Day96CPs}, computed with the
amplitude-fitted prior.  For these we show both times, which exhibit
slightly different distributions, though in both cases the closure
phases are dominated by peaks near $-10^\circ$ 
and $10^\circ$.  The probability of measuring a closure phase within
the allowed range exceeds 97\% at both times.
The closure phase of the model with the highest amplitude-fitted
prior probability (based upon the $1.3\,\mm$-VLBI visibility
amplitude) is located at $12.6^\circ$.  However, significant
probabilities extend out to 
$\Phi\simeq\pm30^\circ$, and thus factor of few improvements in
sensitivity necessarily result in a significant likelihood of
detecting non-trivial closure phases.

\section{Closure Phases on Future Triangles} \label{sec:CPFT}
The development of the EHT will be characterized by the inclusion of
additional baselines and improvements in station sensitivity.  Thus,
we also compute the closure phases along triangles including
intermediate and large north-south baselines.  Figure
\ref{fig:AllCPs}
show the probability distribution for closure phases on all triangles
made with the SMT, CARMA, Hawaii, the LMT, and Chile, as seen at 12:20
UTC on day 96 of 2009.  Closure phase distributions on other days and
times are qualitatively similar.  
While the
closure phase on the SMT-CARMA-Hawaii triangle is small, this is not
generic. 
Closure phases with magnitudes of $\sim45^\circ$ are likely
on five of ten triangles, including the SMT-Hawaii-Chile and
Hawaii-Chile-LMT triangles, for which the most probable model has
closure phases comparable to $-90^\circ$ (shown by the red vertical
lines in each panel)\footnote{Note that in 30\% of the cases shown the
red line lies outside the $1$-$\sigma$ region.  This is not unexpected
given the definition of the $1$-$\sigma$ and $2$-$\sigma$ regions, and
does not indicate that the model with the highest posterior
probability is inherently unlikely.  These deviations are generally
largest for large open triangles, associated with the intrinsically
larger value and distribution of closure phases in these cases.}.  

More importantly, with improvements in the station sensitivity,
arising from increases in bandwidth and the phasing together of
multiple elements at individual stations, substantial improvements in
the precision with which closure phases can be measured are possible.
Despite the small visibilities associated with the most probable model
\citep[see][]{Brod_etal:10}, on all baselines constructed from the
five stations we consider, Sgr A* is detectable on the typical
atmospheric coherence time ($10\,\s$) with SNR exceeding unity.  Over
the dynamical time scale of Sgr A*, roughly $10\,\min$, the
uncertainty in the closure phase is reduced by an additional order of
magnitude, resulting in the red, horizontal error bars in the
individual panels of Figure \ref{fig:AllCPs},
ranging from $0.6^\circ$--$1.8^\circ$.  In all cases this is much smaller
than the presently allowed range of values, and provides sufficient
precision to distinguish likely values from zero.

\section{Improving the Constraints Upon Sgr A*'s Spin}
\begin{figure*}
\begin{center}
\includegraphics[width=\textwidth]{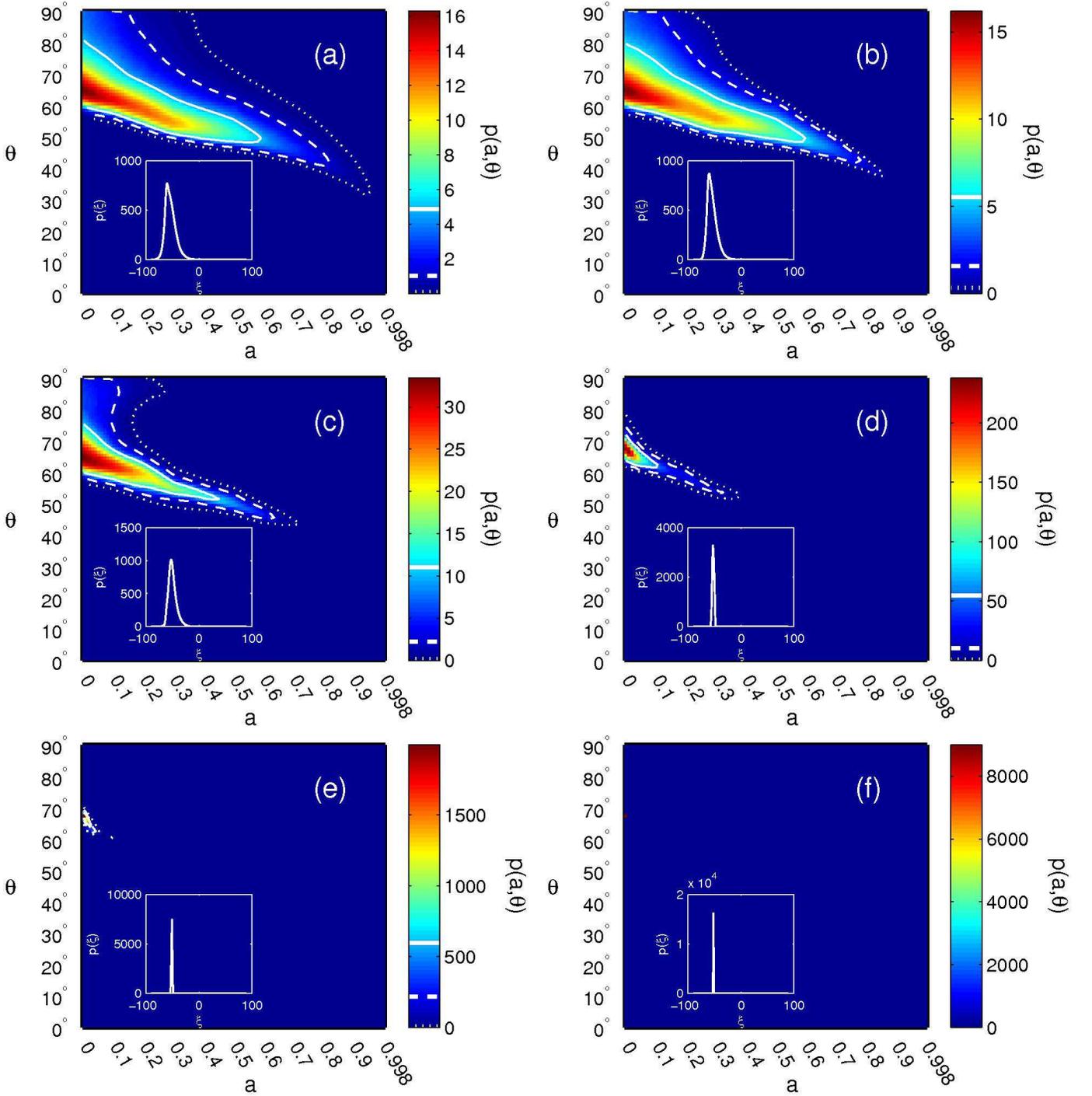}
\end{center}
\caption{Marginalized posterior probability distributions for the spin
magnitude and inclination, and the spin position angle (inset), for
various closure phase measurements.  In particular, shown are the
marginalized posterior probabilities when: (a) only the visibility
amplitudes are considered (note that this is somewhat more broadly
distributed as a result of marginalizing over $\xi$), (b) the closure
phase measurement reported in \citet{Fish_etal:10} is included, (c)
a single closure phase measurement of $12.6^\circ\pm5^\circ$ on the
SMT-CARMA-JCMT triangle is included, (d) closure phases are measured
on all triangles connecting the Hawaii, CARMA, SMT, and APEX
stations, comparable to a single ALMA dish, assuming the
presently most likely model, (e) closure phases are measured on all
triangles connecting the Hawaii, CARMA, SMT, and Chile stations, using
10 ALMA dishes, assuming the presently most likely model, and (f) when
all triangles connecting the Hawaii, CARMA, SMT, Chile and LMT stations
are considered, assuming the most likely model.  In the final
instance, the relevant probability region is no longer resolved,
dominated overwhelmingly by a single pixel.  See the text for more
details.} \label{fig:CPspinlimits}
\end{figure*}

The high precision with which closure phases may be measured in the
near future, warrants a brief examination of their ability to
help constrain the spin of Sgr A*.  To estimate this we consider the
effects of including a single closure phase measurement on each of the
various station-triangles discussed in Section \ref{sec:CPFT}.  In
practice, where many closure phase measurements for each triangle are
anticipated, these are expected to be even more constraining.

Figure \ref{fig:CPspinlimits}a shows the posterior probability
distribution obtained from analysing the 1.3mm-VLBI amplitudes alone,
as a function of spin magnitude ($a$) and inclination ($\theta$),
after marginalized over the spin position angle ($\xi$)\footnote{Note
  that the allowed region is somewhat broader than shown in Figure
  \ref{fig:1220CPs} due to this marginalization.}.  Including the
closure phase measurement reported in \citet{Fish_etal:10} eliminates
some of the high-spin, high-inclination models, producing the
probability distribution seen in Figure \ref{fig:CPspinlimits}b.
Nevertheless, the $1$-$\sigma$ and $2$-$\sigma$ contours are not
significantly affected.  The represents the present constraints upon
the spin of Sgr A*.

To illustrate the improvement associated with higher-precision
measurements and additional triangles, we will assume the most likely
model, as defined by the 1.3mm-VLBI visibility amplitudes alone, is
the ``true'' model for Sgr A*'s emission region.  This allows us to
compute the visibilities throughout the $u$-$v$ plane, obtain the SNRs
at each point, and thus infer both the expected closure phase,
$\Phi_{ijk,0}$ and the precision with which it may be measured,
$\sigma_{\Phi_{ijk,0}}$.  Based upon this, we can construct a
likelihood associated with these observations,
\begin{equation}
\mathcal{L}(\bmath{a}) = C \mathcal{L}_{\rm |V|}(\bmath{a})
\exp\left[-\sum_{ijk}
\frac{\left[\Phi_{ijk,0}-\Phi_{ijk}(\bmath{a})\right]^2}{2\sigma_{\Phi_{ijk,0}}^2}
\right]\,,
\end{equation}
where $C$ is some constant and $\mathcal{L}_{\rm |V|}$ is the
likelihood of observing the measured visibility amplitudes alone.  This may
then be used to produce a posterior probability density,
$p(\bmath{a})$), as described in \citet{Brod_etal:10}, the particulars
of which depend upon which closure phase measurements we choose to
include and their associated measurement uncertainties.  In what follows
we consider a handful of possible measurements that are relevant for
upcoming observations.

Improving the precision of the closure phase measurement on the
SMT-CARMA-JCMT triangle to $\pm5^\circ$ substantially improves the
spin estimation in two respects.  Firstly, it would conclusively
eliminate the $180^\circ$ degeneracy in the position angle resulting
from fitting visibility amplitudes alone.  Secondly, as seen in Figure
\ref{fig:CPspinlimits}c, it begins to significantly modify the
1-$\sigma$ and 2-$\sigma$ contours.  Thus, even with modest
precisions, closure phase measurements can substantial improve black
hole spin estimation in Sgr A*.  

For comparison with ongoing observations, we show in Figure
\ref{fig:CPspinlimits}d the constraints associated with closure phases
measured with a phased Hawaii, phased CARMA, SMT, and APEX, all with a
$1\,\GHz$ bandwidth (in contrast to the $4\,\GHz$ bandwidth assumed
elsewhere).  Even without the substantial sensitivity improvements
associated with large bandwidths and phasing up multiple ALMA
antennas, the resulting constraint is strengthened dramatically, in
principle allowing a spin magnitude estimate with precision $\pm0.1$.

The spin constraints associated with observations with a phased ALMA
are shown in Figure \ref{fig:CPspinlimits}e, at which point the
allowed model parameter space becomes too small to resolve with the
parameter sampling we employed.  When the LMT is included
(Figure \ref{fig:CPspinlimits}f), the
allowed region is reduced to a single grid point in our model space
($\Delta a=0.01$, $\Delta\theta=\Delta\xi=1^\circ$).
Whether such high precision spin estimation is possible in practice is
almost certainly going to be limited by the systematic uncertainties
in the modeling of the emission region.  Nevertheless, the
constraining power of even a handful of closure phase measurements is
clear.

\section{Conclusions}\label{sec:C}
The measurement of mm-VLBI closure phases in Sgr A* provide an
independent, significant constraint upon the structure of the emission
region.
Models based upon radiatively inefficient accretion flows do not
generally have small closure phases, and thus are not generally
consistent with the recently measured value on the SMT-CARMA-JCMT
triangle.  As a consequence, even in the presence of large
uncertainties, closure phases can place significant constraints upon
the structure of Sgr A*.

However, when only 
accretion flow models that fit the $1.3\mm$-VLBI visibility
amplitudes are considered, the probability of finding an excluded
closure phase falls to less than 3\%.  Hence, the vast majority of the
acceptable models in \citet{Brod_etal:10} are broadly consistent with
all current $1.3\mm$-VLBI constraints.
This is surprising since the
$1.3\,\mm$-VLBI amplitudes and closure phase measurements provide
independent constraints, and thus having fit the accretion models to
the former there is no a priori reason to expect that they
would be consistent with the latter.
Based upon the models that are consistent with the $1.3\mm$-VLBI
visibility amplitudes, we predict that the closure phase on the
SMT-CARMA-JCMT triangle during the time it was measured was
between $\pm30^\circ$, with $\pm13^\circ$ most preferred.
Improving the SNR of mm-VLBI observations by a factor of a few,
achievable by phasing up multiple CARMA antennas and employing the
recently phased sites in Hawaii, should allow detection of a non-zero closure
phase on the SMT-CARMA-JCMT triangle.  Note that this would
eliminate the existing $180^\circ$ ambiguity in the orientation of Sgr
A*.

If Sgr A* is well described by radiatively inefficient
accretion flow models, currently acceptable closure phases on large
triangles (e.g., SMT-Hawaii-Chile and Hawaii-Chile-LMT) can be as large as
$90^\circ$, and typical values on triangles which include long
baselines of $45^\circ$.  Future improvements in sensitivity
associated with the near-term development of the EHT will
substantially increase the precision with which closure phases can be
measured.  On all triangles we considered (constructed from the SMT,
CARMA, Hawaii, the LMT, and Chile) the resulting precision should be
sufficient to measure the deviations from $0^\circ$.

Improved closure phase measurements and baseline coverage result in
dramatically improved constraints upon the spin of Sgr A*.  Including
a single antenna in Chile (e.g., APEX or a single ALMA dish) produces
spin estimates with precisions of roughly $\pm0.1$.  With a partially
phased ALMA station, this improves to $\pm0.03$, and with the LMT is
better than $\pm0.01$, at which point we fail to resolve the likely
region.  In the latter cases, the spin estimation will almost certainly
be limited by systematic uncertainties in the modeling of the emission
region motivating the consideration of more sophisticated accretion
and outflow models.

\acknowledgments
This work was supported in part by NSF grants AST-0907890,
AST-0807843 and AST-0905844, and NASA grants NNX08AL43G and
NNA09DB30A.

\bibliographystyle{apj}

\begin{thebibliography}{99}
\expandafter\ifx\csname natexlab\endcsname\relax\def\natexlab#1{#1}\fi

\bibitem[{{Bower} {et~al.}(2006){Bower}, {Goss}, {Falcke}, {Backer}, \&
  {Lithwick}}]{Bowe_etal:06}
{Bower}, G.~C., {Goss}, W.~M., {Falcke}, H., {Backer}, D.~C., \& {Lithwick}, Y.
  2006, \apjl, 648, L127

\bibitem[{{Broderick} \& {Blandford}(2004)}]{Brod-Blan:04}
{Broderick}, A., \& {Blandford}, R. 2004, \mnras, 349, 994

\bibitem[{{Broderick}(2006)}]{Brod:06}
{Broderick}, A.~E. 2006, \mnras, 366, L10

\bibitem[{{Broderick} {et~al.}(2009){Broderick}, {Fish}, {Doeleman}, \&
  {Loeb}}]{Brod_etal:09}
{Broderick}, A.~E., {Fish}, V.~L., {Doeleman}, S.~S., \& {Loeb}, A. 2009, \apj,
  697, 45

\bibitem[{{Broderick} {et~al.}(2010){Broderick}, {Fish}, {Doeleman}, \&
  {Loeb}}]{Brod_etal:10}
---. 2010, ArXiv:1011.2770

\bibitem[{{Broderick} \& {Loeb}(2006{\natexlab{a}})}]{Brod-Loeb:06a}
{Broderick}, A.~E., \& {Loeb}, A. 2006{\natexlab{a}}, \apjl, 636, L109

\bibitem[{{Broderick} \& {Loeb}(2006{\natexlab{b}})}]{Brod-Loeb:06b}
---. 2006{\natexlab{b}}, \mnras, 367, 905

\bibitem[{{Dexter} {et~al.}(2010){Dexter}, {Agol}, {Fragile}, \&
  {McKinney}}]{Dext-Agol-Frag-McKi:10}
{Dexter}, J., {Agol}, E., {Fragile}, P.~C., \& {McKinney}, J.~C. 2010, \apj,
  717, 1092

\bibitem[{{Doeleman} {et~al.}(2009{\natexlab{a}}){Doeleman}, {Agol}, {Backer},
  {Baganoff}, {Bower}, {Broderick}, {Fabian}, {Fish}, {Gammie}, {Ho}, {Honman},
  {Krichbaum}, {Loeb}, {Marrone}, {Reid}, {Rogers}, {Shapiro}, {Strittmatter},
  {Tilanus}, {Weintroub}, {Whitney}, {Wright}, \& {Ziurys}}]{Doel_etal:2010}
{Doeleman}, S., {et~al.} 2009{\natexlab{a}}, in Astronomy, Vol. 2010,
  astro2010: The Astronomy and Astrophysics Decadal Survey, 68

\bibitem[{{Doeleman} {et~al.}(2009{\natexlab{b}}){Doeleman}, {Fish},
  {Broderick}, {Loeb}, \& {Rogers}}]{Doel_etal:09}
{Doeleman}, S.~S., {Fish}, V.~L., {Broderick}, A.~E., {Loeb}, A., \& {Rogers},
  A.~E.~E. 2009{\natexlab{b}}, \apj, 695, 59

\bibitem[{{Doeleman} {et~al.}(2008){Doeleman}, {Weintroub}, {Rogers},
  {Plambeck}, {Freund}, {Tilanus}, {Friberg}, {Ziurys}, {Moran}, {Corey},
  {Young}, {Smythe}, {Titus}, {Marrone}, {Cappallo}, {Bock}, {Bower},
  {Chamberlin}, {Davis}, {Krichbaum}, {Lamb}, {Maness}, {Niell}, {Roy},
  {Strittmatter}, {Werthimer}, {Whitney}, \& {Woody}}]{Doel_etal:08}
{Doeleman}, S.~S., {et~al.} 2008, \nat, 455, 78

\bibitem[{{Fish} {et~al.}(2011){Fish}, {Doeleman}, {Beaudoin}, {Blundell},
  {Bolin}, {Bower}, {Chamberlin}, {Freund}, {Friberg}, {Gurwell}, {Honma},
  {Inoue}, {Krichbaum}, {Lamb}, {Marrone}, {Moran}, {Oyama}, {Plambeck},
  {Primiani}, {Rogers}, {Smythe}, {SooHoo}, {Strittmatter}, {Tilanus}, {Titus},
  {Weintroub}, {Wright}, {Woody}, {Young}, \& {Ziurys}}]{Fish_etal:10}
{Fish}, V.~L., {et~al.} 2011, \apjl, 727, L36

\bibitem[{{Ghez} {et~al.}(2008){Ghez}, {Salim}, {Weinberg}, {Lu}, {Do}, {Dunn},
  {Matthews}, {Morris}, {Yelda}, {Becklin}, {Kremenek}, {Milosavljevic}, \&
  {Naiman}}]{Ghez_etal:08}
{Ghez}, A.~M., {et~al.} 2008, \apj, 689, 1044

\bibitem[{{Gillessen} {et~al.}(2009{\natexlab{a}}){Gillessen}, {Eisenhauer},
  {Fritz}, {Bartko}, {Dodds-Eden}, {Pfuhl}, {Ott}, \& {Genzel}}]{Gill_etal:09b}
{Gillessen}, S., {Eisenhauer}, F., {Fritz}, T.~K., {Bartko}, H., {Dodds-Eden},
  K., {Pfuhl}, O., {Ott}, T., \& {Genzel}, R. 2009{\natexlab{a}}, \apjl, 707,
  L114

\bibitem[{{Gillessen} {et~al.}(2009{\natexlab{b}}){Gillessen}, {Eisenhauer},
  {Trippe}, {Alexander}, {Genzel}, {Martins}, \& {Ott}}]{Gill_etal:09a}
{Gillessen}, S., {Eisenhauer}, F., {Trippe}, S., {Alexander}, T., {Genzel}, R.,
  {Martins}, F., \& {Ott}, T. 2009{\natexlab{b}}, \apj, 692, 1075

\bibitem[{{Huang} {et~al.}(2009){Huang}, {Takahashi}, \&
  {Shen}}]{Huan-Taka-Shen:09}
{Huang}, L., {Takahashi}, R., \& {Shen}, Z. 2009, \apj, 706, 960

\bibitem[{{Jennison}(1958)}]{Jenn:58}
{Jennison}, R.~C. 1958, \mnras, 118, 276

\bibitem[{{Jennison} \& {Latham}(1959)}]{Jenn-Lath:59}
{Jennison}, R.~C., \& {Latham}, V. 1959, \mnras, 119, 174

\bibitem[{{Jones} \& {O'Dell}(1977)}]{Jone-ODel:77}
{Jones}, T.~W., \& {O'Dell}, S.~L. 1977, \apj, 214, 522

\bibitem[{{Pearson} \& {Readhead}(1984)}]{Pear-Read:84}
{Pearson}, T.~J., \& {Readhead}, A.~C.~S. 1984, \araa, 22, 97

\bibitem[{{Petrosian} \& {McTiernan}(1983)}]{Petr-McTi:83}
{Petrosian}, V., \& {McTiernan}, J.~M. 1983, Physics of Fluids, 26, 3023

\bibitem[{{Rogers} {et~al.}(1995){Rogers}, {Doeleman}, \&
  {Moran}}]{Roge-Doel-Mora:95}
{Rogers}, A.~E.~E., {Doeleman}, S.~S., \& {Moran}, J.~M. 1995, \aj, 109, 1391

\bibitem[{{Rogers} {et~al.}(1984){Rogers}, {Moffet}, {Backer}, \&
  {Moran}}]{Roge_etal:84}
{Rogers}, A.~E.~E., {Moffet}, A.~T., {Backer}, D.~C., \& {Moran}, J.~M. 1984,
  Radio Science, 19, 1552

\bibitem[{{Rogers} {et~al.}(1974){Rogers}, {Hinteregger}, {Whitney},
  {Counselman}, {Shapiro}, {Wittels}, {Klemperer}, {Warnock}, {Clark}, \&
  {Hutton}}]{Roge_etal:74}
{Rogers}, A.~E.~E., {et~al.} 1974, \apj, 193, 293

\bibitem[{{Thompson} {et~al.}(2001){Thompson}, {Moran}, \&
  {Swenson}}]{Thom-Mora-Swen:01}
{Thompson}, A.~R., {Moran}, J.~M., \& {Swenson}, Jr., G.~W. 2001,
  {Interferometry and Synthesis in Radio Astronomy} (2nd ed.; Weinheim:
  WILEY-VCH)

\bibitem[{{Weintroub}(2008)}]{Wein:08}
{Weintroub}, J. 2008, Journal of Physics Conference Series, 131, 012047

\bibitem[{{Yuan} {et~al.}(2003){Yuan}, {Quataert}, \&
  {Narayan}}]{Yuan-Quat-Nara:03}
{Yuan}, F., {Quataert}, E., \& {Narayan}, R. 2003, \apj, 598, 301

\end{thebibliography}

\end{document}